\documentclass[10pt,letterpaper]{article}
\usepackage{opex3}
\usepackage{epstopdf}
\usepackage{graphicx}
\usepackage{cite}

\begin{document}


\title{Efficient single-spatial-mode periodically-poled KTiOPO$_4$ waveguide source for high-dimensional entanglement-based quantum key distribution}


\author{Tian Zhong,$^{1,*}$ Franco N. C. Wong,$^1$
\\Alessandro Restelli,$^2$ and Joshua C. Bienfang$^2$}

\address{
$^1$Research Laboratory of Electronics, Massachusetts Institute of Technology, Cambridge, Massachusetts 02139, USA
\\
$^2$Joint Quantum Institute, University of Maryland and National Institute of Standards and Technology, Gaithersburg, Maryland 20899, USA \\
}

\email{$^*$tzhong@mit.edu}

\begin{abstract}We demonstrate generation of high-purity photon pairs at 1560 nm in a single spatial mode from a periodically-poled KTiOPO$_4$ (PPKTP) waveguide. With nearly lossless spectral filtering, the PPKTP waveguide source shows approximately 80 \% single-mode fiber coupling efficiency and is well suited for high-dimensional time-energy entanglement-based quantum key distribution. Using high-count-rate self-differencing InGaAs single-photon avalanche photodiodes configured with either square or sinusoidal gating, we achieve $>$ 1 Mbit/s raw key generation with 3 bits-per-photon encoding, and, to the best of our knowledge, the highest reported Franson quantum-interference visibility of 98.2 \% without subtraction of accidental coincidences.\end{abstract}

\ocis{(030.5260) Photon Counting; (190.4410) Nonlinear optics, parametric processes;  (270.5570) Quantum detectors; (270.5565) Quantum communications; (270.5585) Quantum information and processing.}

\section{Introduction}
\noindent Conventional quantum key distribution (QKD) uses a discrete two-dimensional Hilbert space for key encoding, such as the polarization state of a single photon. In contrast, high-dimensional QKD allows encoding onto a larger state space, such as multiple levels of a continuous variable of a single photon, thus enabling the system to achieve higher photon information efficiency (bits per photon) and potentially higher key rate (bits per second). However, its deployment requires the development of high-performance source, detector, and routing technologies tailored to the specific large-alphabet encoding scheme.  One such high-dimensional QKD system of interest is based on time-energy entanglement, in which keys are derived from the arrival times of photon pairs generated from continuous-wave (cw) spontaneous parametric downconversion (SPDC). With synchronized clocks, Alice and Bob measure the arrival times of a pair of SPDC-generated temporally coincident photons within a frame composed of 2$^n$ time bins, and extract $n$ bit values \cite{Irfan, Irfan2}.  The maximum number of time bins per frame for key encoding is given by the ratio of the time-frame duration and the two-photon correlation time. For a typical photon-pair source with a correlation time on the order of 1 ps, as many as 20 bits per photon pair can, in principle, be encoded in a 1 $\mu$s frame duration, assuming detectors have timing resolution that is comparable to the SPDC correlation time \cite{Irfan}. 
 
High-dimensional encoding in entanglement-based QKD is particularly advantageous when the photon-pair generation rate $R_{\rm s}$ is significantly lower than the maximum rate supported by the detectors, 1/$\tau_{\rm d}$, where $\tau_{\rm d}$ is the detector timing resolution. In this case, if traditional binary encoding is used, then the maximum the is $R_{\rm s}$ (assuming lossless transmission and detection), and any two detection events are, on average, separated by a large number of empty time bins. In contrast, with the same source generation rate $R_{\rm s}$, high-dimenstional encoding across a frame of length 2$^n \approx (R_{\rm s} \cdot \tau_{\rm d})^{-1}$ yields $n$ bits per detection event, increasing the maximum throughput rate to $nR_{\rm s}$. With typical source generation rates in the 10$^6$ s$^{-1}$ range and detector resolution of the order of 50 ps, this benefit can be sizeable. In such a scenario, high-dimensional encoding makes more efficient use of the available detection resources and can significantly increase throughput rates.  
 
Implementation of a complete high-dimensional QKD system requires various hardware (source and detector) and software (suitable protocol with proven security, error-correction coding, and privacy amplification) components. The goal of this work is to design and demonstrate a capable entanglement source with compatible detector technology as a first step towards a functional high-speed, high-dimensional QKD system. Prior theoretical works \cite{Boure, Cerf} suggest two critical hardware requirements for optimal performance in high-dimensional QKD: first, the source and detectors should have high overall efficiency; second, Alice and Bob should measure a maximum correlation between their photons throughout QKD operation. Using a pulsed SPDC source and avalanche photodiode single-photon detectors, Marcikic {\it et al.} demonstrated distribution of time-bin entangled photons (with binary encoding) at telecom band over 50 km of standard single-mode fibers \cite{marc}. However, that early system suffered from low coincidence counts due to fiber losses, limited detector efficiency, and a limited pulsed-source repetition rate of 75 MHz, and therefore had low key rates. Subsequent works used sources based on periodically-poled lithium niobate (PPLN) waveguides to boost pair generation rates and fiber coupling efficiency \cite{takesue, zhang, dynes}. Despite their high brightness,  broadband PPLN waveguide SPDC sources may incur substantial losses when narrowband spectral filtering is applied to limit the output bandwidth, resulting in less than optimal coincidence rates. In addition, the entanglement quality of time-bin qubits, often used in fiber-optic systems, has not achieved the same level as that of the more commonly used polarization qubits; while polarization entanglement with $\geq$ 99 \% quantum-interference visibilities is routinely obtained \cite{sagnac,Fedrizzi}, the highest reported raw Franson interference \cite{Franson1989} visibility is only 95.6 \% \cite{Tittel2010}. 

In this paper we present a source of time-energy entangled photons that addresses many of the issues critical to the implementaiton of multi-dimensional QKD at high rates. We have developed a periodically-poled KTiOPO$_4$ (PPKTP) waveguide source of entangled pairs in a single spatial mode that uses nearly-lossless wideband spectral filtering and exhibits a generation efficiency of $1.0\times10^{7}$ pairs/(s$\cdot$nm$\cdot$mW)\@.  Using high count-rate self-differencing (SD) InGaAs single-photon avalanche diodes (SPADs) \cite{shields, alessandro, alessandro2} configured with square or sinusoidal gating at 628.5 MHz, we have experimentally demonstrated $>$ 1 Mbit/s raw key generation with 3 bits-per-photon encoding, and achieved, as a measure of entanglement quality, a 98.2 \% Franson interference visibility without subtraction of accidental coincidences. 

The paper is organized as follows. Section 2 details the properties of the waveguide source and shows how we take advantage of the spectral properties of higher-order spatial modes to create a single-spatial-mode SPDC output. Section 3 describes the high-count-rate self-differencing InGaAs SPAD detection system, which is critical for achieving the performance we observe. The main results of the time-energy entangled-photon source are given in Section 4, and we conclude with a discussion in Section 5.
 
\section{Single-spatial-mode waveguide source of time-energy entangled photons}
Many applications in quantum information science benefit from compact sources of entangled photons that are efficient and bright, and waveguided SPDC sources meet these requirements nicely. Moreover, in fiber-based quantum communications, the ability to easily select a single spatial mode and couple efficiently into a SMF are important factors that can significantly affect the entanglement quality and key generation. In this respect, waveguide sources have a distinct advantage over bulk-crystal sources because it is relatively easy to remove higher-order spatial modes from waveguide sources with minimal filtering loss. 

The transverse confinement in a waveguide imposes on the two-photon emission a discrete set of spatial modes defined by the waveguide's intrinsic modal dispersion properties \cite{fiorentino, eckstein, zhong, mosley}. The spectral separation between different spatial modes can be engineered to be large enough that a wideband bandpass filter (BPF) is sufficient to select one and only one spatial mode. We note that wideband BPFs with high transmission efficiency are commercially available, and the spectral filtering does not require very steep edges as long as the flat-top pass band covers the entire two-photon phase-matching bandwidth.  Consequently, for a waveguided SPDC source, single-mode output is experimentally achievable with nearly-lossless spectral filtering. In contrast, bulk-crystal SPDC emits into a continuum of spatial modes with overlapping but not quite identical spectral content. Therefore it is difficult to efficiently isolate a specific spatial mode for coupling into a SMF\@. It is common to apply narrowband filtering (with interference filters or Bragg-grating filters) to obtain a well defined spectrum that also removes some undesirable spatial modes. In this case, however, the narrowband filter incurs extra losses unless it has very steep edges. 

We have developed a photon-pair source based on a 15.6 mm long PPKTP waveguide \cite{advr} with a cross section (width $\times$ height) of 4 $\mu$m $\times$ 8 $\mu$m that supports multiple spatial modes at telecom wavelengths. The 46.1 $\mu$m grating period was designed for type-II quasi-phase-matched wavelength-degenerate SPDC at 1560 nm in the fundamental modes of the signal and idler fields. The pump was a 780 nm Ti:Sapphire laser coupled into the fundamental mode of the waveguide with 75 \% input coupling efficiency. We directly imaged the SPDC outputs using an ultrasensitive InGaAs short-wave infrared camera \cite{camera}, as shown in Fig.~1(a). The main spot of the imaged output that is framed by the dashed box is the fundamental waveguide mode. The large cloud of light surrounding the main spot originated from higher-order spatial modes in the waveguided SPDC process. Using the models developed in \cite{zhong, fiorentino}, our simulations show that these higher-order modes are associated with significantly different sets of wavelengths: the SPDC spectra of the first few lowest order modes (other than the fundamental mode) are at least 70 nm away from the fundamental-mode wavelength of 1560 nm. Therefore, we were able to apply a 10 nm band-pass filter with 99 \% transmission to spectrally remove the higher-order spatial modes and to transmit only the desired photon pairs with negligible losses. Figure~1(b) displays the image of the spectrally filtered SPDC, clearly showing the nearly-circular signal and idler fundamental modes. 

\begin{figure}[htb]
\label{fig1}
\centering
\includegraphics[width=5in]{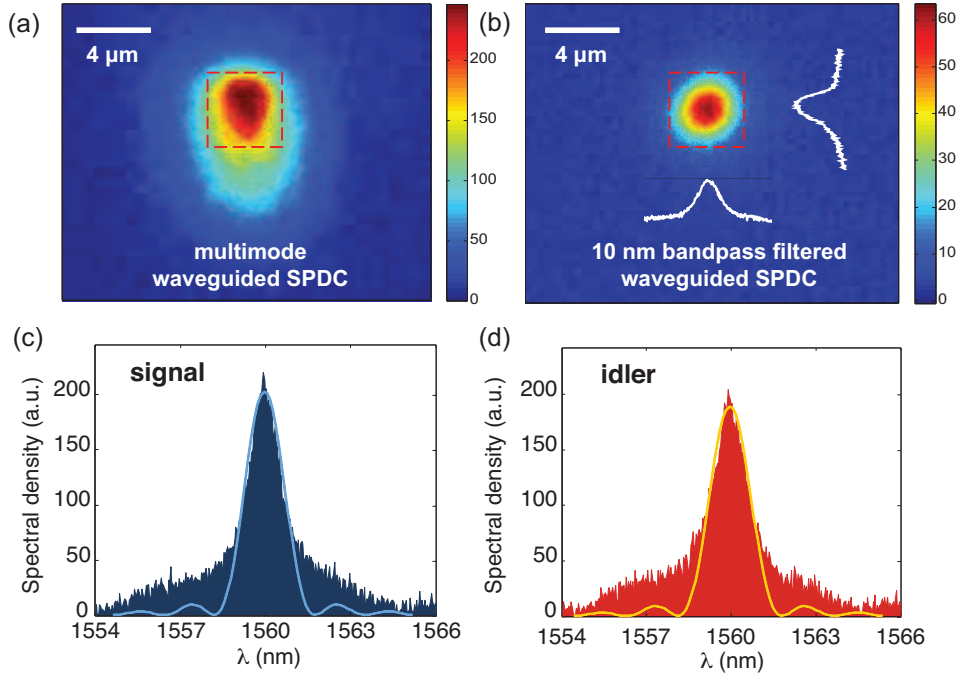}
\caption{Spatial and spectral properties of waveguide SPDC. (a) Multimode waveguided SPDC output. Red dashes outline the approximate area of the fundamental mode of 4 $\mu$m $\times$ 4 $\mu$m, which covers a 30 $\times$ 30 pixel area on the InGaAs infrared camera. (b) Fundamental-mode waveguided SPDC after 10 nm band-pass spectral filtering, with transverse mode profiles. Optical spectrum of signal (c), and idler (d), after 10 nm band-pass filtered and coupled into a single-mode fiber. Both are fitted using $sinc^2$ functions. The signal and idler phase-matching bandwidth is 1.6 nm.}
\end{figure}

After spectral filtering we were able to couple the signal and idler beams into polarization-maintaining SMF by optimizing the mode matching using a custom built zoom-lens. We achieved high waveguide-to-fiber coupling efficiencies of 79 \% and 85 \% for the signal and idler fundamental modes, respectively. These coupling efficiencies were first measured using a probe laser carefully aligned through the waveguide in the signal and idler fundamental modes. The values were subsequently verified by comparing the spectrally filtered SPDC power collected from a multimode fiber to that from a SMF\@. Overall, the symmetric conditional spectral-spatial collection efficiency is approximately 80 \%, which is among the highest ever reported for a waveguide entanglement source.

We should point out that the SMF also served as a spatial filter to remove any residual high-order spatial modes that potentially exist within the 10 nm bandwidth of the BPF\@. To confirm the absence of higher-order modes, we measured the optical spectra of the signal and idler photons using a diffractive-grating-based single-photon spectrometer (similar to the one used in \cite{mosley}). The results with background subtracted are plotted in Fig.~1(c) and (d), showing a single spectral peak of the fundamental spatial mode, which suggests strong temporal/spectral correlations of the fiber-coupled photon pairs. Such single-spatial-mode operation is crucial for achieving a high visibility in Franson quantum interference, because any unintentionally collected spurious modes will lower the conditional coupling efficiency and cause an increased background accidental coincidences. We also fit the measured spectrum using a $sinc^2$ function. The pedestals in Fig.~1(c) and (d) are most likely due to non-uniformity of the periodic grating structure, which has been reported previously for similar PPKTP waveguides \cite {zhong, fiorentino}.  Lastly, we measured a spectral brightness of the fundamental mode (at the source) of approximately 10$^7$ pairs/(s$\cdot$nm$\cdot$mW) of pump, and a signal and idler phase-matching bandwidth of 1.6 nm, in good agreement with the theoretical estimate.   

\section{High count-rate self-differencing InGaAs SPADs}
Recently, systems that apply periodic bias gates to InGaAs SPADs at rates in the gigahertz range have been shown to enable high-speed counting ($>$ 10$^8$ s$^{-1}$) of telecom-wavelength photons using commercially available SPADs \cite{shields, namekata}. In these systems, the short duration of the bias gate, as well as its periodic nature, allow the discrimination of otherwise undetectably small avalanches from signals produced by the gates themselves. We previously constructed a high-speed periodically-gating InGaAs detection system based on the self-differencing (SD) technique \cite{shields}, that is capable of operating with both square-wave and sinusoidal gating waveforms \cite{alessandro}. When using a square-wave waveform we can achieve good detection efficiency (roughly 20 \%) in a short gate. With a sinusoidal waveform in the SD system we achieve better transient cancellation, which gives us enhanced sensitivity to smaller avalanches, resulting in a longer effective gate width and higher detection duty cycle, while also reducing the total amount of charge necessary for discrimination, which reduces the afterpulse probability \cite{namekata, jzhang}.
  
We characterized the SD InGaAs SPADs using correlated photon pairs from our waveguide source by measuring the coincidences using two SPADs that were synchronously gated with either square-wave or sinusoidal waveforms, at 628.5 MHz. For both waveforms the over voltage was about 3 V, corresponding to a detection efficiency of roughly 20 \% at 1560 nm. The gate duration was 500 ps for the square-wave gate and 900 ps for the sinusoidal gate. By illuminating the SPADs at $−20^{\circ}$C with a flux of 0.1 photons per gate once every 16 periods and counting the total number of avalanches in the illuminated and non-illuminated gates (as in \cite{alessandro2}), we measured an after-pulse probability of 6.0 \% (integrated over 15 gates) for square-wave gates and 3.5 \% for sinusoidal gates, with a hold-off time of 80 ns imposed by the recovery time of the counting electronics. Figure 2 shows the coincidence results as the relative delay between the two SPADs was scanned using an electronic phase shifter. As can be seen in Fig. 2, the improved transient cancellation achievable with the sinusoidal waveform results in a significantly wider detection window than that can be achieved with a square wave, resulting in a detection duty cycle of 25 \% for sinusoidal gating, and 7 \% for square gating. The difference in duty cycle makes the two gating configurations suitable for different types of measurements; the high duty cycle achievable with sinusoidal gating is ideal for high-coincidence-rate detection, while the narrower square-wave gating is desirable for high-visibility quantum interference measurements because it provides better discrimination against multi-pair events and background counts. Here the multi-pair events refer to independent, Poisson-distributed  temporal modes of SPDC generated in one detection window.

\begin{figure}[htb]
\label{fig2}
\centering
\includegraphics[width=3in]{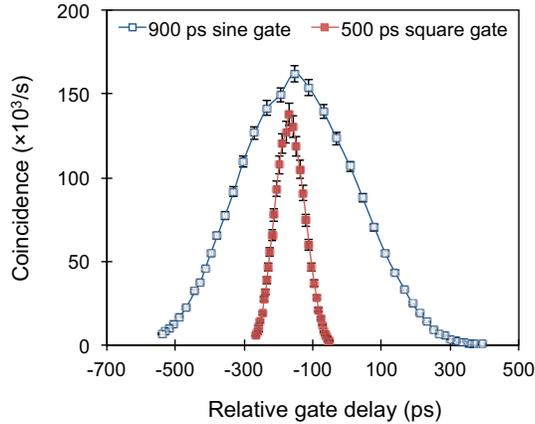}
\caption{Detected coincidences of correlated photon pairs versus the relative gate delay between two SPADs with 500 ps square gating (filled square) and 900 ps sinusoidal gating (open square). After deconvolution, the effective gate widths are found to be 110 ps and 395 ps for square gating and sinusoidal gating, respectively. The error bars are based on 5 \% estimated measurement uncertainty. }
\end{figure}

\section{High rate measurements of time-energy entangled photons}
Figure 3 shows the experimental setup for the high rate time-energy entanglement measurements. After spectral filtering and fiber coupling, the orthogonally polarized photon pairs from the waveguide source were separated into signal and idler using a fiber polarizing beam splitter (PBS). We characterize the performance of our source using the time-energy entanglement-based QKD setup in \cite{Irfan2} with two pairs of SPADs, one with sinusoidal gating for key generation, and the other with square-wave gating for monitoring security in a Franson interferometer \cite{Irfan, Irfan2}. An optical switch was used to manually choose between the two measurements. Following the upper branch of the switch for key generation, we operated the InGaAs SPADs using sine gating with its wider effective gate window at a gate frequency of either 628.5 MHz (1.6 ns time bin) or 1.257 GHz (0.8 ns time bin). In the former case, the detection efficiencies were 22 \% and 20 \% with a dark count probability of $<$ 3$\times$10$^{-5}$per gate for the two SPADs. In the latter case, the detection efficiencies were 18 \% and 17 \% and the dark count probability was $<$ 2$\times$10$^{-5}$per gate. Each detection event by Alice and Bob was time tagged and recorded by a time-to-digital converter (Hydraharp 400). Synchronization between Alice and Bob's detections was achieved by characterizing the delay between the signal and idler channels and determining the time-binned coincidences from the recorded events. Every time-binned coincidence then yields a key of $n$ bits information when the time frame duration is $2^n$ time bins. In this way, multiple bits per photon were extracted using time-bin encoding with a frame duration of 6.4 ns, corresponding to 2 bits per photon when gating the detectors at 628.5 MHz, and 3 bits per photon at 1.257 GHz gating.

\begin{figure}[htb]
\label{fig3}
\centering
\includegraphics[width=\textwidth]{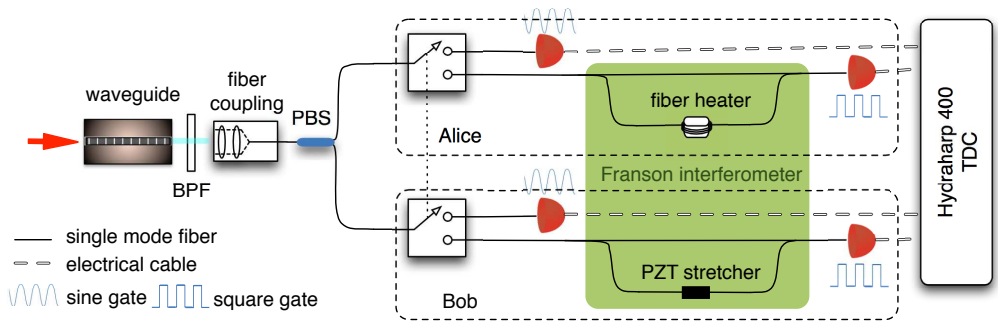}
\caption{Schematic of experimental setup. Upper branch of the switch is for key generation where the SPADs used sinusoidal gating for higher coincidence rates. Lower branch of the switch is for Franson interferometry where narrow square-wave gating was used for better visibility. The Franson interferometer consists of two unbalanced Mach-Zehnder interferometers, which are constructed using 50:50 fiber beam splitters. BPF: band-pass filter, PBS: polarizing beam splitter, TDC: time-to-digital converter.}
\end{figure}

The maximum overall system efficiencies for signal and idler were 11.2 \%  and 9.6 \%, respectively, with 628.5 MHz sine-gated InGaAs SPADs. Besides the band-pass filtering and the fiber coupling, other losses include 0.65 dB/cm (1.35 dB/cm) waveguide internal attenuation of the signal (idler), 0.65 dB (0.75 dB) insertion loss of the fiber PBS for signal (idler), and 0.5 dB coating loss for various free-space optical components. The internal attenuation of the waveguide was measured using the Fabry-Perot interferometry technique in \cite{regener}. Since photon pairs were generated with equal probability along the entire length of the waveguide, we take the average internal loss to be half of the measured attenuation coefficients. Overall, thanks to the high source efficiency as well as 25 \% detection duty cycle, the detected singles count rate on each channel was close to 10$^7$ s$^{-1}$ with less than 50 mW of pump power. At such high rates, the count saturation is dominantly caused by the 80 ns recovery time of each time-to-digital converter (TDC) and not by the detector dead time, which, with the SD technique, is much less than 80 ns. To avoid miscounting due to counter saturation, we electrically switched each detector's output into two TDC channels in an alternating fashion. Doing so effectively reduced the counter dead time from 80 ns per channel to 40 ns, allowing multi-million events per second to be recorded faithfully. Figure 4 plots the accidentals-subtracted coincidence rates versus the pump power. At a pump power of 35 mW, we recorded 7.73 $\times$ 10$^5$ s$^{-1}$ and 5.86 $\times$ 10$^5$ s$^{-1}$ coincidences for 628.5 MHz and 1.257 GHz gating rate, respectively, corresponding to raw key rates of 1.55 Mbit/s with 2 temporal bits per photon, or 1.76 Mbit/s with 3 bits per photon using the same 6.4 ns time frame. 

The accidental coincidences at both gate frequencies were also recorded and displayed in Fig. 4 (open squares and open circles). Multi-pair generation was the dominant source of accidentals, whereas the contribution due to dark counts was insigniﬁcant in this high rate measurement. Detector jitter makes no contribution to the accidentals rate because the gate duration is shorter than the coincidence measurement window. Being dominated by multi-pair generation, we observe that the accidental coincidences increase quadratically with pump power. In the context of high-dimensional QKD with time-energy entanglement, accidental coincidences would be the dominant cause of frames in which Alice and Bob each have a detection event, but in different time bins within the frame. Such non-coincident events in recorded frames generate errors in the raw key. However, the nature of high-dimensional encoding, as well as the use of Franson interferometry as a security check, complicate the impact such errors have on both the number of bit errors in the raw key and the relation between the observed bit error rate and the information gain of an eavesdropper. The measurement of accidentals alone does not directly characterize the error-corrected and privacy-amplified performance of high-dimensional QKD; this relationship is an area of active research.

\begin{figure}[htb]
\label{fig4}
\centering
\includegraphics[width=3in]{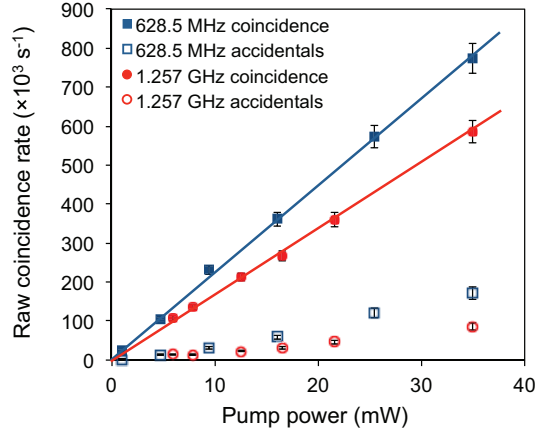}
\caption{Measured coincidence rates at different pump powers using InGaAs SPADs sinusoidally gated at 628.5 MHz and 1.257 GHz. Accidental coincidences (open squares and open circles) are used to obtain the accidentals-subtracted rates (solid squares and solid circles). Straight lines show the linear extrapolation. The error bars are based on one standard deviation.}
\end{figure}

The lower branch of the switch in Fig. 3 leads to an all-fiber Franson interferometer for evaluating the quality of time-energy entanglement, which in a complete QKD implementation could provide security against an eavesdropper's positive-operator valued measure (POVM) attack \cite{Irfan2}. The SD InGaAs SPADs were operated with square gating at 628.5 MHz to produce a narrow coincidence window of roughly 100 ps at full-width-half-maximum. A narrow gating window reduces the probability of multi-pair events per gate, which is a common factor in degrading entanglement quality and therefore the Franson interference visibility. In a real QKD implementation, however, this narrow gating window has to sample the entire key-generation gate at some point in time, either by scanning or some other means, so that no gap between the two gating configurations can be exploited by the eavesdropper. Using the square gating we measured dark count rates of $<$ 2$\times$10$^{-6}$ per gate for the two SPADs, which were significantly lower than the photon count rates throughout our measurements at various pump powers, suggesting that the accidental coincidences due to detector dark counts were negligibly small. 

Each arm of the interferometer was enclosed in a thermally sealed box for improved phase stability. The path mismatch in each arm of the interferometer was 4.8 ns, and the difference in the two path mismatches was set to zero within the two-photon coherence time of about 2 ps (0.3 mm in fiber) by using bright 75 ps pulses and ultrafast photodetection of their rising edges through the two arms. Once set, the zero path-mismatch difference (within 0.3 mm) could be maintained for hours. In addition, for fine tuning of the path-mismatch difference, we used both a closed-loop temperature control of Alice's long-path fiber and a piezoelectric transducer (PZT) fiber stretcher on Bob's long path. 

\begin{figure}[htb]
\label{fig5}
\centering
\includegraphics[width=\textwidth]{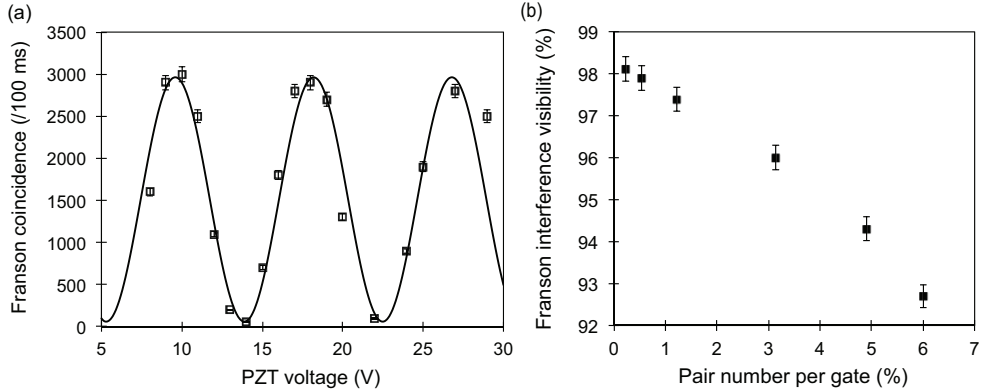}
\caption{(a) Franson quantum interference fringes measured with narrow square gating at pair generation number per gate $\alpha$ = 3.1\%. (b) Measured Franson visibilities (without subtraction of accidentals) versus $\alpha$. The error bars are based on one standard deviation.}
\end{figure}

Figure 5(a) shows the Franson interference fringes by scanning the PZT fiber stretcher on Bob's long path.  At 35 mW pump power, the pair generation number per gate was $\alpha$ = 3.1 \%, and we measured a Franson visibility $V = (C_{\rm max}-C_{\rm min})/ (C_{\rm max}+C_{\rm min})$ of 96.0 $\pm$ 0.3 \%  without subtraction of accidentals, where $C_{\rm max}$ and $C_{\rm min}$ are the maximum and minimum measured coincidences, respectively. We repeated the visibility measurements at different pump powers, as plotted in Fig. 5(b). The degradation of the Franson visibility with increasing pump power shows good agreement with the expected $V = 1 -\alpha$ relationship when multi-pair events are taken into account \cite{marc2}. More importantly, we observed the highest visibility of 98.2 $\pm$ 0.3 \% without accidentals subtraction at a mean pair number of $\alpha=0.24$ \%. Besides the degradation due to multi-pair events (0.24 \%), two other factors might have prevented us from obtaining even higher visibility. One is the accidental coincidences that we measured to be 0.2 \% which were not excluded in the raw visibility measurements. The second factor is the dispersion due to the 4.8 ns path-length difference in each arm of the Franson interferometer. For a standard fiber, the dispersion coefficient of 17 ps/(nm$\cdot$km) should yield a temporal spread of 0.026 ps which is about 1 \% of the photon pulse width. The roughly 1 \% estimated dispersion mismatch in the pulse widths is compatible with the observed visibility degradation. Dispersion effects in Franson interferometry will be discussed in more details in a separate publication. The present results are better than previously reported visibility measurements even with their background subtraction. The high quality of time-energy entanglement and its maximum spectral/temporal correlations between signal and idler of our waveguide source were made possible by the high purity single-spatial-mode operation of our waveguide source.

\section{Conclusion}
In conclusion, the key to achieving high secure bit rates in high-dimensional entanglement-based QKD is to operate the source with high efficiency and high entanglement quality. Compatible measurement capabilities must be available to handle the high count rates and to confirm the high entanglement quality as a security measure. We have developed a source of single-spatial-mode photon pairs at 1560 nm based on a PPKTP waveguide with nearly-lossless spectral filtering and approximately 80 \% fiber coupling. The waveguide source is fundamentally superior to bulk-crystal sources because of the non-overlapping nature of the waveguide's spectral-spatial mode structure. Using high count-rate self-differencing InGaAs SPADs with either square or sinusoidal gating, we have measured raw key generation exceeding 1 Mbit/s with up to 3 bits per photon, and we have achieved the highest Franson quantum-interference visibility to date of $>$ 98 \% without subtraction of accidentals. The simultaneous achievements of high count rate and high quality quantum interference represent a significant improvement over previous systems. With proper error correction and privacy amplification, our system can be readily applied to implementing high dimensional QKD with improved secret key rates.

\section*{Acknowledgments}
This work was supported by the DARPA InPho program, including US Army Research Office award W911NF-10-1-0416.


\begin{thebibliography}{99}
\bibitem{advr} Manufactured by AdvR Inc. The identification of any commercial product or trade name does not imply endorsement or recommendation by the National Institute of Standards and Technology.
\bibitem{camera} Goodrich SU320HX-1.7RT high sensitivity InGaAs SWIR camera. The identification of any commercial product or trade name does not imply endorsement or recommendation by the National Institute of Standards and Technology.
\end{thebibliography}
\end{document}